\documentclass[prl,letterpaper,twocolumn,showpacs,amsmath,amsfonts,amssymb,preprintnumbers]{revtex4}
\pdfoutput=1
\usepackage[T1]{fontenc}\usepackage[latin1]{inputenc}
\usepackage{dcolumn,graphicx,color,hvfloat,booktabs,microtype,afterpage} \graphicspath{{Figures/}}
\renewcommand{\figurename}{Fig.}
\renewcommand{\tablename}{Table}
\makeatletter\renewcommand{\fnum@figure}[1]{\figurename~\thefigure.}\makeatother
\makeatletter\renewcommand{\fnum@table}[1]{\tablename~\thetable.}\makeatother

\usepackage[colorlinks,plainpages=false,linkcolor=blue,urlcolor=blue,citecolor=black,pdfpagemode=UseNone,pdfstartview=FitBH]{hyperref}

\begin{document}

\title{LaFeAsO$_{1-x}$F$_{x}$ thin films: high upper critical fields and evidence of weak link behavior.}

\author{S.~Haindl{\large\hyperref[CorrAuthor]{*}}}
\author{M.~Kidszun}\author{A.~Kauffmann}\author{K.~Nenkov}\author{N.~Kozlova}\author{J.~Freudenberger}\author{T.~Thersleff}\author{J.~Hänisch} 
\author{J.~Werner}
\author{E.~Reich}\author{L.~Schultz}\author{B.~Holzapfel}
\affiliation{IFW Dresden, P.\,O.\,Box 270116, D--01171 Dresden, Germany.}

\begin{abstract}
Superconducting LaFeAsO$_{1-x}$F$_{x}$ thin films were grown on single crystalline LaAlO$_{3}$ substrates with critical temperatures (onset) up to 28\,K. Resistive measurements in high magnetic fields up to 40\,T reveal a paramagnetically limited upper critical field, $\mu_{0}H_{c2}$(0) around 77\,T and a remarkable steep slope of $-6.2$\,TK$^{-1}$ near $T_{c}$. From transport measurements we observed a weak link behavior in low magnetic fields and the evidence for a broad reversible regime.  
\end{abstract}

\pacs{\vspace{-0.2em}74.25.Dw 74.25.Qt 74.25.Sv 74.78.Bz\vspace{-0.5em}}

\maketitle
Since the discovery \cite{01, 02, 03} of high temperature superconductivity in the \textit{R}FeAsO$_{1-x}$F$_{x}$ (\textit{R} = rare earth) iron pnictides, the so called `1111' phase or family, and subsequently in the intermetallic Ba$_{1-x}$K$_{x}$Fe$_{2}$As$_{2}$, the `122' phase, an intensive investigation of this material class started \cite{04,05} and related compounds of the FeSe and the LiFeAs--structure, respectively the `11' and `111' phase, have been found \cite{06,07}. The quaternary iron pnictides exhibit high superconducting transition temperatures up to 55\,K, high upper critical fields, and a multiband character. Consequently their classification between MgB$_{2}$ and the cuprates is heavily discussed to date. This sandwich position and the vicinity of magnetic order \cite{08} makes them a good candidate not only for new phenomena in superconductivity but also for an understanding of the pairing mechanism in high temperature superconductors including the cuprates.

Similarities between the cuprate high temperature superconductors and the iron pnictides have already been pointed out, and the analogy is mainly based upon the layered structure, the charge carrier doping, and the vicinity of an antiferromagnetic phase \cite{04,09}. On the other hand, there are crucial differences between the cuprates and the iron pnictides. Very recent theoretical results on the basis of tight--binding calculations highlight the multi--band character of the electronic band structure, and a clear two--dimensional behavior of the band structure near the Fermi energy was found for the `1111' phase \cite{10}. In addition, a possible Pauli limitation of the upper critical field is still under discussion \cite{20,21}.

In case of the `1111' phase most of the experimental investigation were undertaken using polycrystals and the available NdFeAsO$_{1-x}$F$_{x}$ and SmFeAsO$_{1-x}$F$_{x}$ single crystals \cite{11,12}. The first superconducting LaFeAsO$_{1-x}$F$_{x}$ thin film has been grown on LaAlO$_3$ substrate by pulsed laser deposition (PLD) \cite{13}. Other thin films are reported for the `122' phase in Ref.\,\cite{14} and for the `11' structure in Ref.\,\cite{15}. PLD is regarded as a versatile tool for thin film fabrication \cite{16}, and one of its advantages is the stoichiometric transfer of target material to the substrate. However, it seems to remain a difficult task to obtain superconductivity in the case of the iron based fluorine doped oxypnictide thin films. A crucial point is the stoichiometric control of the fluorine content in the grown samples and poses a challenge to the growth of the quaternary compounds. In addition, due to the high reactivity of the rare earth elements (especially La), the suppression of oxide phases is a key issue for thin film deposition. We succeeded in the fabrication of a LaFeAsO$_{1-x}$F$_{x}$ thin film with optimal critical temperature and a complete superconducting transition. High--quality thin films are mandatory for electronic devices, multilayers or Josephson junctions from the new superconductors, which may lead to new effects based on the interplay of superconductivity and magnetism. Detailed results of superconducting LaFeAsO$_{1-x}$F$_x$ single crystals are not reported so far, therefore the deposition of thin films is of enormous interest for fundamental studies. For instance, the behavior of grain boundaries is still an open question, whether they act as pinning sites or as weak links.

We present the first detailed transport investigation of superconducting LaFeAsO$_{1-x}$F$_{x}$ thin films in magnetic fields up to 40\,T. Our measurements demonstrate a high upper critical field and Pauli limitation. In our transport measurements we observe small critical current densities and weak link behavior in this iron pnictide superconductors.

\begin{figure}[t]
	\centering
		\includegraphics[width=\columnwidth]{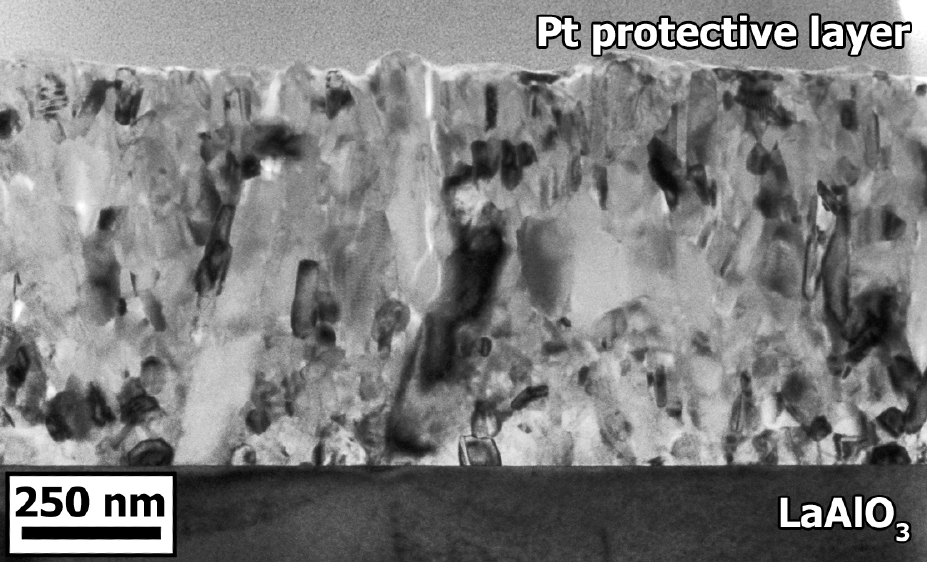}
		\caption{Bright field TEM image (orientation contrast) of the LaFeAsO$_{1-x}$F$_x$ thin film. The film is polycrystalline with a significant portion of grains growing elongated perpendicular to the substrate surface. No pores were observed.} 
\label{fig:figure1}
\end{figure}

\begin{figure}[t]
	\centering
		\includegraphics[width=0.7\columnwidth]{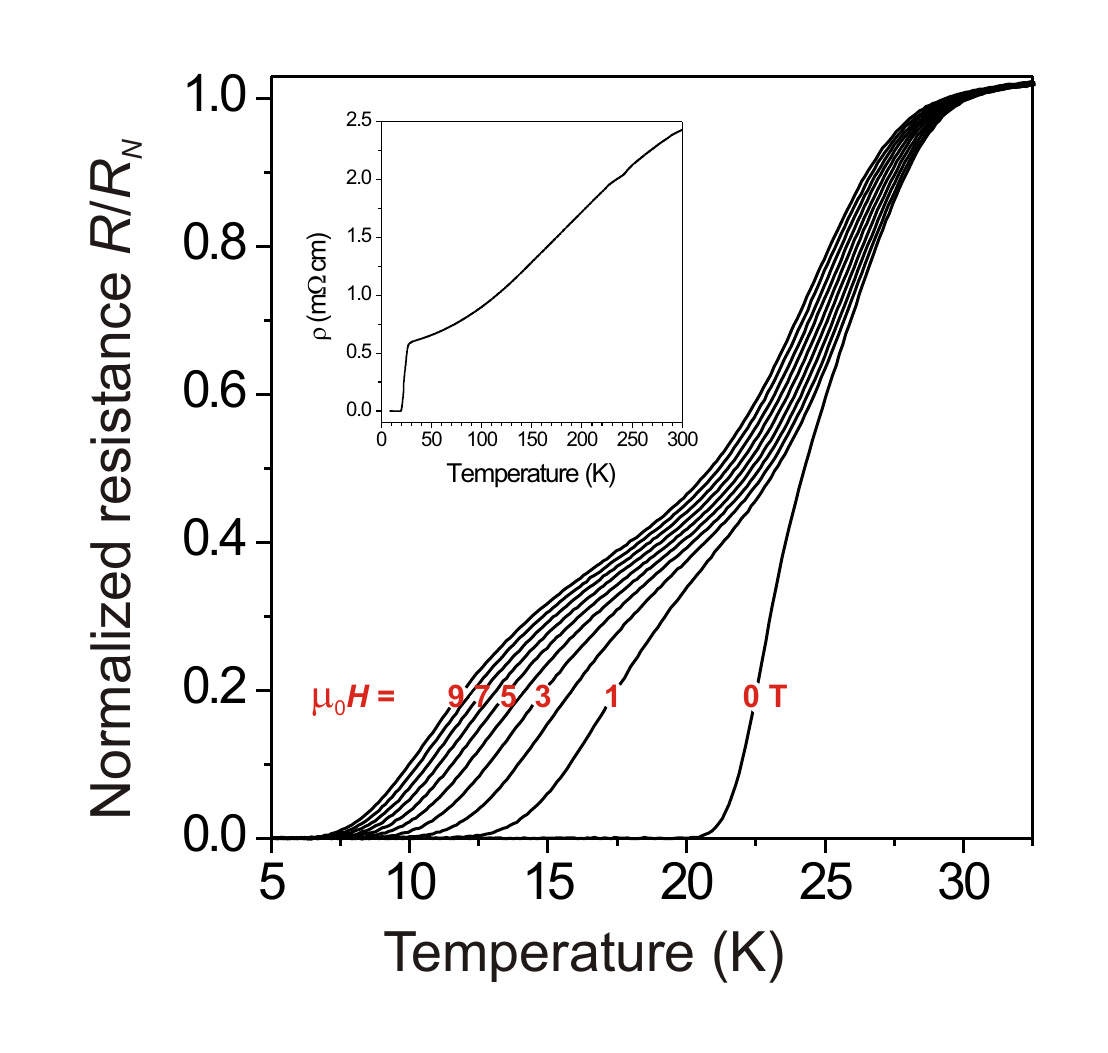}
		\caption{The resistivity shows metallic behavior and superconductivity below a temperature of 28\,K with a heavy broadening of the superconductive transition in an external magnetic field (shown up to 9\,T).} 
\label{fig:figure2}
\end{figure}

Thin films were grown at room temperature on single crystalline LaAlO$_{3}$(001) (LAO) substrates by standard on--axis PLD using a KrF laser (Lambda Physik) with a wavelength of $\lambda = 248$\,nm, a pulse duration of $\tau = 30$\,ns, and an energy density of $\epsilon \approx 4$\,Jcm$^{-2}$ at the target surface. Vacuum conditions in the deposition chamber were $p_{\text{base}} = 10^{-6}$\,mbar. The nominal target composition is LaFeAsO$_{0.75}$F$_{0.25}$. An \textit{ex--situ} post annealing step at 950$\:^{\circ}$C for 4\,hours in a sealed quartz tube followed the room temperature deposition. Thin film preparation details can be found in Ref.\,\cite{13}. A further reduction of the oxygen partial pressure, compared to previous experiments, results in a complete superconducting transition to zero resistivity and reduces the fraction of impurity phases (LaOF, La$_{2}$O$_{3}$). Structural investigation by transmission electron microscopy (TEM) reveals a homogeneous film without pores. The cross section bright field image (Fig.\,\ref{fig:figure1}), where the orientation contrast dominates, in combination with Fast Fourier Transformation (FFT) shows polycrystallinity and columnar grains elongated perpendicular to the substrate surface. From detailed electron energy loss spectroscopy (EELS) as well as energy dispersive x--ray analysis (STEM--EDX) we conclude phase homogeneity in the superconducting layer examined by TEM.  

The `1111' phase was identified by X--ray diffraction (Bragg--Brentano) using Co K$_{\alpha}$ radiation showing (003), (110) and (200) as the three strongest reflections. The lattice constants derived are $a = 4.02$\,\r{A} and $c = 8.65$\,\r{A}. A cut piece with dimensions of 1\,mm$\times$5\,mm with a film thickness of 700\,nm, $T_{c,90} = 28$\,K and residual resistivity ratio (RRR) of 4.1 (inset in Fig.\,\ref{fig:figure2}) was used for resistive measurements. No spin density anomaly was found around 150\,K, and from the comparison with the temperature dependence of the resistivity in polycrystals for different fluorine doping levels, a fluorine content of above $\sim 10\%$ is estimated. Four probe electrical transport measurements were carried out in a commercial Physical Property Measurement System (PPMS) up to magnetic fields of 9\,T (100\,$\mu A$ dc current) as well as up to 14\,T (100\,$\mu A$ ac amplitude). Both $R(T)$ and $R(\mu_{0}H)$ data have been recorded in static fields.

\begin{figure}[t]
	\centering
		\includegraphics[width=1.15\columnwidth]{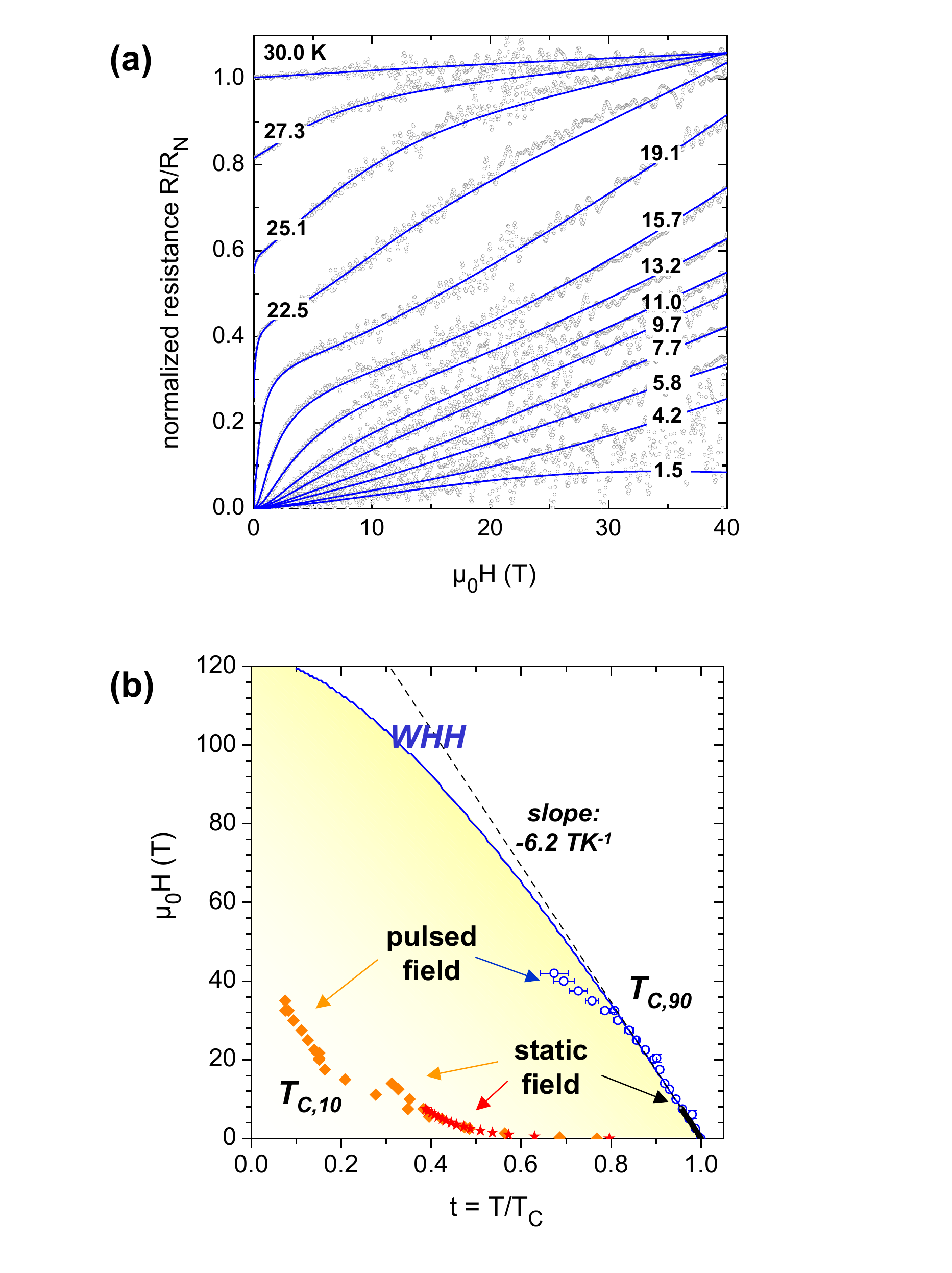}
		\caption{(a) The pulsed field data (circles) and corresponding flattened curves are shown up to 40\,T for different temperatures. (b) The magnetic phase diagram of the LaFeAsO$_{1-x}$F$_{x}$ thin film. $T_{c,90}$ and $T_{c,10}$ of measurements in pulsed fields and static fields are given. The low field $T_{c,90}$ from PPMS measurements fit the pulsed field data well and exhibit a slope near $T_{c}$ of $-6.2$\, TK$^{-1}$ (dashed line). The upper critical field data deviates from the WHH model.}
	\label{fig:figure3}
\end{figure}

Resistive measurements, $R(T)$, show well defined and fully developed transitions from the normal to the superconducting state with $R = 0$ (Fig.\,\ref{fig:figure2}). The broadening of the superconductive transitions in applied magnetic fields can be ascribed to the effects of \textit{i}) a polycrystalline structure of the thin film, and \textit{ii}) low critical current densities. Pulsed field measurements, $R(\mu_{0}H)$, up to 40\,T were performed in a cryostat equipped with a solenoid, which is operated in pulsed mode (with an ac amplitude of 95 $\mu$A and a frequency of 10\,kHz). A detailed description of the pulsed field system can be found elsewhere \cite{18}. Heating of the sample during the pulse is negligible, as checked by several test pulses with successively increasing maximum field. All measurements were carried out for an orientation of the film surface perpendicular to the magnetic field direction.   

\begin{figure}[t]
	\centering
		\includegraphics[width=0.95\columnwidth]{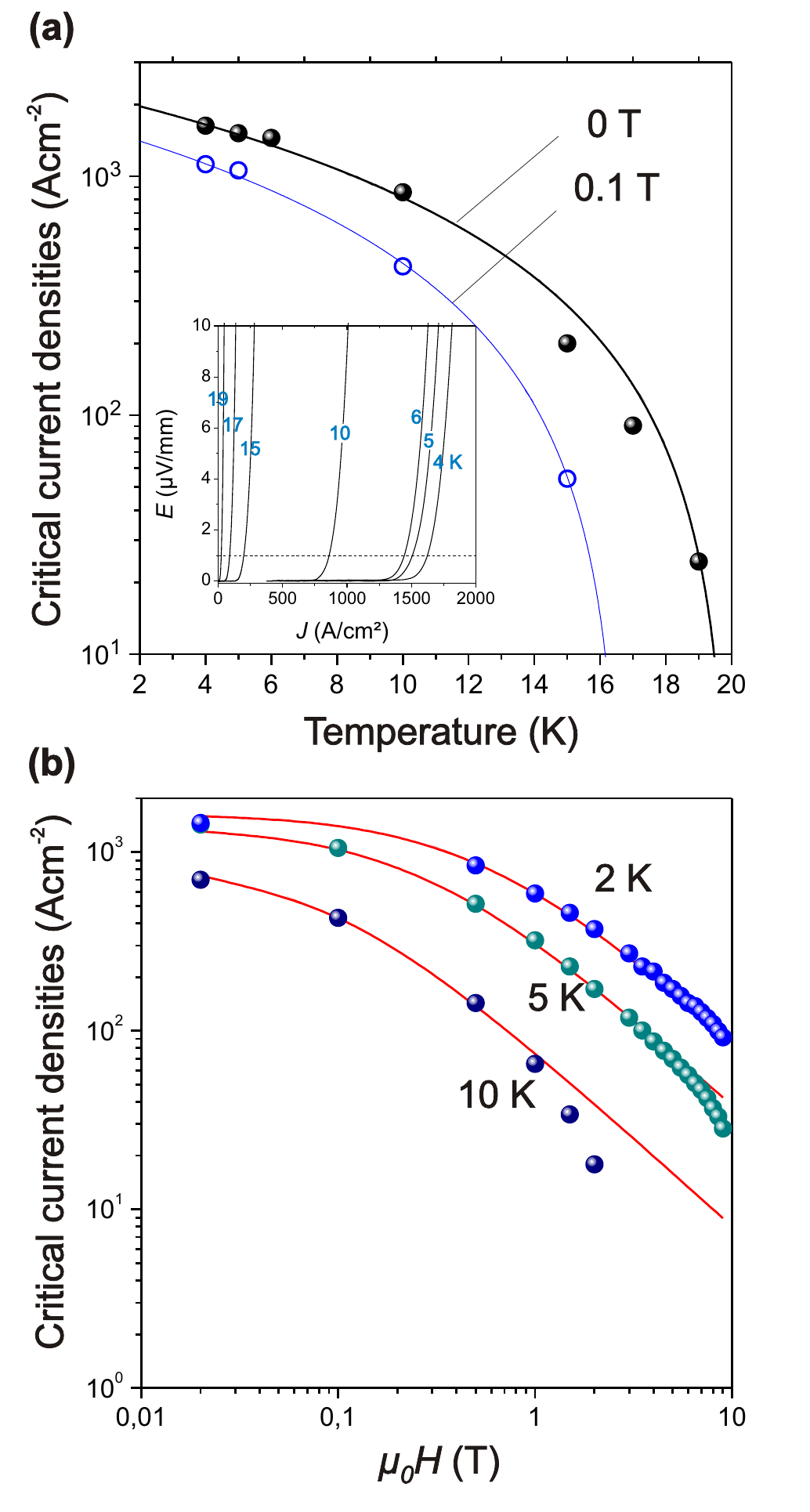}
		\caption{(a) Temperature dependence of the critical current densities in zero field ($\bullet$) and in a small applied field of 0.1\,T ($\circ$). The inset shows the $E(J)$--measurements in zero field for different temperatures (4, 5, 6, 10, 15, 17 and 19\,K) and the 1\,$\mu$Vmm$^{-1}$ criterion for evaluation.(b) Field dependence of the critical current densities for temperatures of 2 K, 5 K and 10 K. The fits (lines) correspond to the behavior of weakly linked grains.}
	\label{fig:figure4}
\end{figure}

The resistive data of the pulsed field measurements for different temperatures show a linear increase in an extended field interval (Fig.\,\ref{fig:figure3}\,(a)). There is a qualitative agreement between the pulsed field data and the low field $R(\mu_{0}H)$ data from the PPMS measurements up to 14\,T, in $T_{c,90}$. Deviations between low--field (PPMS) and pulsed field measurements can be observed for the evaluation of $T_{c,10}$ at low temperatures. The collapse of the current density at higher fields and temperatures, addressed further below, can explain the observed behavior well if one takes into account that the measurement current of 100\,$\mu$A exceeds the critical current. The additional perturbation of possibly pinned flux by the application of an \textit{ac} current in the pulsed field measurements versus a \textit{dc} current in the PPMS measurements further supports this interpretation. The upper critical field obtained by the $T_{c,90}$ criterion (Fig.\,\ref{fig:figure3}\,(b)) shows a remarkable slope near $T_{c}$ of $\vert \frac{d\mu_{0}H_{c2}}{dT} \vert_{T_{c}} = 6.2$\,TK$^{-1}$. This high value exceeds the previously ones for LaFeAsO$_{1-x}$F$_{x}$ \cite{20,21} and is at the lower limit of the values found in NdFeAsO$_{1-x}$F$_{x}$ single crystals \cite{09,22}. The Werthamer--Helfand--Hohenberg (WHH) estimation of the upper critical field at zero temperature yields $0.693 T_{c} \vert \frac{d\mu_{0}H_{c2}}{dT} \vert_{T_{c}} = 120$\,T \cite{23}. The deviation from the WHH model with increasing applied magnetic fields indicates a Pauli limited upper critical field which can be estimated with a coupling constant $\lambda = 0.5$ to around $\mu_{0}H^P = (1+\lambda) \mu_{0}H^{P}_{\text{BCS}} = 77$\,T, where $\mu_{0}H^{P}_{\text{BCS}} = 1.84 T_{c}$ is the BCS paramagnetic limit (for weak coupling).  

Transport current measurements for zero field and $\mu_{0}H = 0.1$\,T evaluated for an electrical field criterion of 1\,$\mu$Vmm$^{-1}$ (Fig.\,\ref{fig:figure4}\,(a)) show a fast decrease of the critical current density, $J_{c}$, with increasing temperature. The temperature dependence of $J_{c}$ is proportional to $(1-t)^{1.5}$, with $t = T/T_{c,10}$ where $T_{c,10}$ is taken from the offset of the resistive transition. The observed small values of $J_{c}$ at low temperatures in the thin film (< 2\,kAcm$^{-2}$) are comparable with the results in LaFeAsO$_{0.9}$F$_{0.1}$ powder--in--tube (PIT) wires \cite{24}. There are several possible explanations to the small absolute values of the critical current densities, and the polycrystalline structure of the film is a strong candidate for the critical current limitation. Experiments on bicristalline YBa$_{2}$Cu$_{3}$O$_{7-\delta}$ thin films have demonstrated that high--angle grain boundaries are able to reduce the critical current by a factor of 10$^{3}$ \cite{25}. Although there is no pronounced weak--link behavior seen in the $E(J)$--curves (see inset in Fig.\,\ref{fig:figure4}\,(a)), the field dependence of the critical current densities (Fig.\,\ref{fig:figure4}\,(b)) suggests strongly a weak link behavior due to grain boundaries (compare Fig.\,\ref{fig:figure1}) parallel to the applied field direction. The fit to the experimental data follows well the description of weakly linked grains with $J_{c} \propto \big(1+\frac{\mu_{0}H}{B^{\star}}\big)^{-1}$ with $B^{\star}$ being a characteristic field of the weak links \cite{17}. In addition, a linear relation between voltage and current has been observed at increased magnetic fields and temperatures, which indicates a dominant reversible regime in the magnetic phase diagram. The irreversibility or depinning line is normally estimated by the offset of the critical temperature in $R(T)$ measurements (Fig.\,\ref{fig:figure2}), but detailed magnetization measurements have to be made in order to confirm this interpretation in the discussed thin films. Since the nature of the vortex matter in the quaternary iron pnictides is still unclear, and the possibility of vortex pancakes, for example, due to the alternating stacking of FeAs and LaO$_{1-x}$F$_{x}$ layers cannot be excluded at the moment, further investigation of flux pinning are of high interest. The problem of granularity and the enormously complex analysis of a vortex dynamical regime has been pointed out in Ref.\,\cite{26} for polycrystals. Therefore, a difference in inter--grain and intra--grain vortex dynamics can also be expected in the polycrystalline iron pnictide thin films. The dissipative signature in the resistive transitions points towards a strong similarity with Bi$_{2}$Sr$_{2}$CaCu$_{2}$O$_{8}$ (Bi--2212) or other highly two--dimensional cuprate superconductors \cite{27}, although the mass anisotropy in the oxypnictides is definitive lower (for instance, $\gamma \approx 5-9$ in Ref.\,\cite{22}). In the first instance, this is quite intriguing, but completely reasonable with regard to the two--dimensional confinement of the charge carriers within the FeAs layer. Certainly, the investigation of the irreversible properties and the vortex matter in the new iron pnictides will be a necessary task in order to fully understand the magnetic phase diagram of these materials. The grain boundaries in the investigated thin film do not play the role of effective pinning centers since the coherence length obtained from the Ginzburg--Landau relation, $\mu_{0}H_{c2} = \frac{\Phi_{0}}{2\pi\xi^{2}}$, is only 2.1\,nm. Consequently, the increase of the critical current densities in the new superconducting iron pnictides depends drastically on the possibility of the introduction of pinning sites into the material as well as on the reduction of the number of high--angle grain boundaries.   

To conclude, we have demonstrated the growth of superconducting LaFeAsO$_{1-x}$F$_{x}$ thin films with a critical temperature of 28\,K and a steep increase ($-6.2$\,TK$^{-1}$) of the upper critical field. The Pauli limited upper critical field, $\mu_{0}H_{c2}(0)$, was estimated to $\sim 77$\,T in accordance with experimental data on polycrystalline bulk material. From transport current measurements there is evidence of a broad dissipative (reversible) regime. The field dependence of the critical current densities support the fact of a weakly linked network in accordance to the very short coherence length, $\xi$, similar to the cuprates. Due to the fact that grain--boundaries in this material act as weak links epitaxial thin films will be necessary for fundamental experimental investigation, including phase sensitive tests \cite{28}, the fabrication of Josephson junctions and multilayers with new interesting properties.
   
\textit{Acknowledgments.} 
The authors would like to thank Marco Langer, Ulrike Besold, Margitta Deutschmann and especially Stefan Pofahl for technical support and Tetyana Shapoval for critical reading of the manuscript. We acknowledge S. Baunack, C. Deneke and O. G. Schmidt for support with the preparation of the TEM sample.

\end{document}